%% file: iclr2027_conference.tex
\documentclass{article} % For LaTeX2e
\usepackage{iclr2027_conference,times}

\input{math_commands.tex}

\usepackage{hyperref}
\usepackage{url}

\usepackage{amsmath}
\usepackage{mathtools}
\usepackage{amsthm}
\usepackage{cuted}
\usepackage{amsfonts}       % blackboard math symbols
\usepackage{nicefrac}       % compact symbols for 1/2, etc.
\usepackage{microtype}      % microtypography
\usepackage{graphicx} 
\usepackage{booktabs}
\usepackage{multirow}
\usepackage[most]{tcolorbox}
\usepackage{xcolor}
\usepackage{xspace}
\usepackage{bm}
\usepackage{enumitem}
\usepackage{pifont}
\usepackage{longtable}
\usepackage{multirow} % 合并单元格
\usepackage[table]{xcolor} % 单元格着色
\usepackage{caption} % 标题控制
\usepackage{adjustbox} % 表格缩放
\usepackage{subcaption} % 必须添加这个宏包以支持subtable
\usepackage{hyperref}
\newcommand{\eg}{\textit{e.g.},\xspace}
\newcommand{\ie}{\textit{i.e.},\xspace}

\title{Auto-Bidding with Disentangled Advertiser Profiles and Train-Free Adaptation}

\author{
Songyue Cai$^{1,2\ddagger*}$ \quad
Shan Gu$^{2\ddagger}$ \quad
Wei Chen$^2$ \quad
Ziru Xu$^2$ \quad
Lianyu Wang$^2$ \quad
Jian Xu$^{2\dagger}$  \\
\textbf{Xiaofeng Zhu}$^{3\dagger}$ \quad
\\
$^1$University of Electronic Science and Technology of China, Chengdu, China \\
$^2$Taobao \& Tmall Group, Alibaba, Beijing, China \\
$^3$Hainan University, Haikou, China \\
\\
\texttt{sonnycai4649@gmail.com,} \\
\texttt{\{gushan.gs, ganhai.cw, ziru.xzr, wanglianyu.wly, xiyu.xj\}@taobao.com,} \\
\texttt{zhuxf@hainanu.edu.cn}
}

\makeatletter
\def\blfootnote{\gdef\@thefnmark{}\@footnotetext}
\makeatother

\iclrfinalcopy
\begin{document}

\maketitle
\blfootnote{$^{*}$Work done during an internship at Taobao \& Tmall Group, Alibaba.}
\blfootnote{$^{\dagger}$Corresponding authors.}
\blfootnote{$^{\ddagger}$Equal contribution}

\begin{abstract}
Auto-bidding is a key component of modern advertising systems that provides a personalized bidding strategy for each advertiser. By characterizing each individual, profile-based methods achieve personalization and have proven effective in domains such as recommendation. However, despite the diverse bidding behavior of advertisers, their application to auto-bidding remains limited. A primary reason is that constructing and leveraging advertiser profiles face several challenges: extracting pure profiles is non-trivial, modeling common and private information simultaneously is difficult, and profile updating and cold-start adaptation remain challenging. To tackle these issues, we propose \textbf{ADAPT}, an \underline{\textbf{A}}uto-bidding framework with \underline{\textbf{D}}isentangled \underline{\textbf{A}}dvertiser \underline{\textbf{P}}rofiles and \underline{\textbf{T}}raining-free adaptation. ADAPT introduces a two-stage training paradigm and supports training-free adaptation. Specifically, (i) the stage 1 extracts pure static and dynamic profiles via contrastive learning over the advertiser memory bank; (ii) the stage 2 disentangles the dynamic profile into a common profile and a private profile, and combines them with the static profile to jointly condition the bidding strategy; (iii) once trained, ADAPT constructs profiles for new advertisers and updates profiles of existing advertisers without retraining. Our experiments on a large-scale auto-bidding benchmark demonstrate that ADAPT consistently achieves superior performance, and ablation studies further validate the effectiveness of each module.The source code will be released at \textcolor[rgb]{0.7, 0.0, 0.125}{\url{https://github.com/YuzunoKawori/ADAPT}}.
\end{abstract}

\section{Introduction}
In online advertising systems, advertisers aim to secure billions of daily impression opportunities through bidding, yet the massive traffic volume and fierce competition~\cite{wine2009internet} make manual bidding by human experts infeasible~\cite{evans2009online,wang2015real}. Auto-bidding addresses this challenge by generating precise bids for large-scale traffic in real time from the current state and environmental information. It formulates bidding as a constrained sequential decision-making problem, dynamically adjusting the bid throughout the delivery period to maximize conversions under advertiser-specified constraints such as cost-per-action (CPA)~\cite{he2021unified,mou2022sustainable}. This makes auto-bidding a crucial mechanism for advertisers to achieve their business objectives in competitive markets~\cite{ren2017bidding}.

\begin{figure*}[t!] % [t] 表示将图片浮动到页面顶部
\centering 
\includegraphics[width=\textwidth]{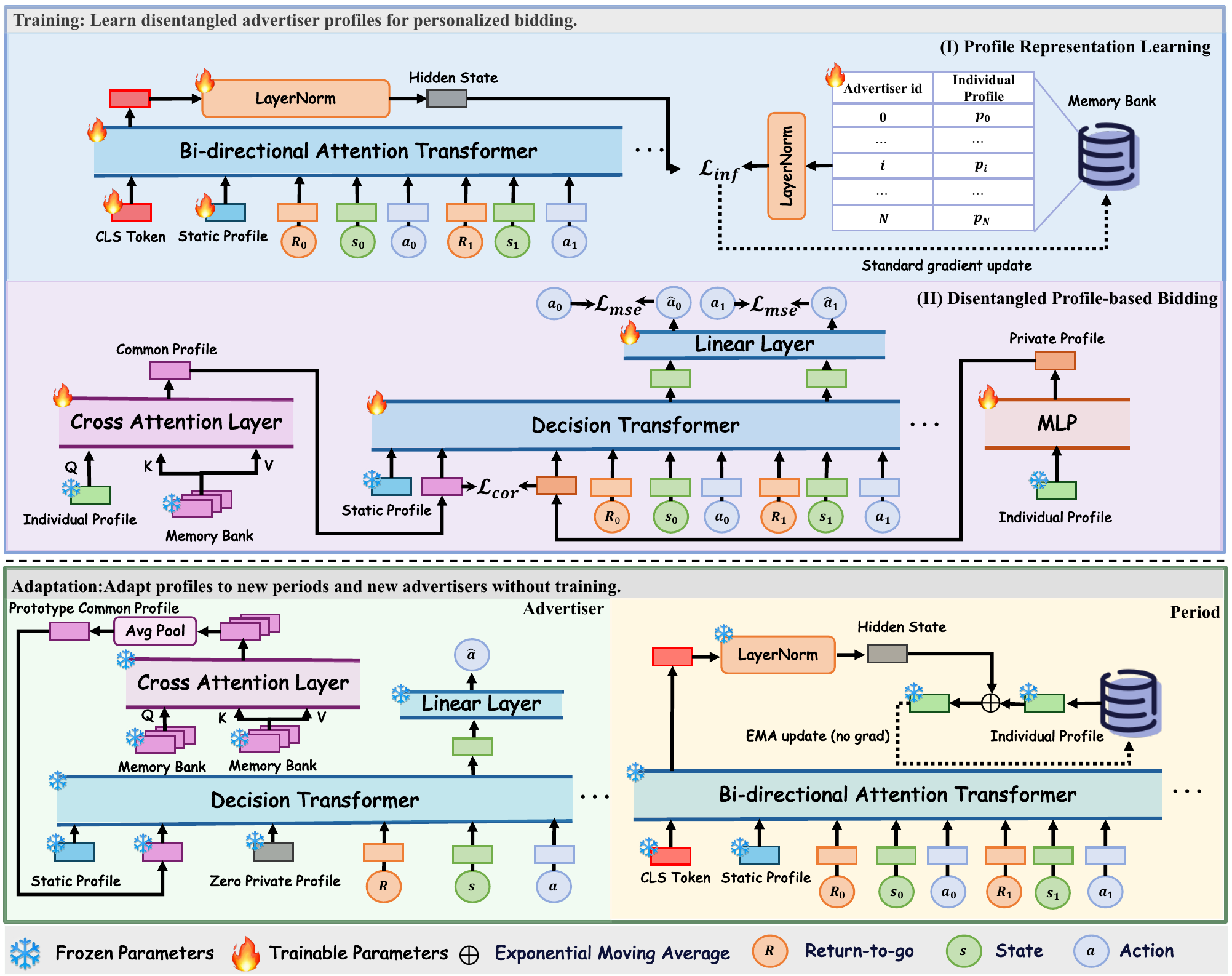} 
\vspace{-1.4em} 
\caption{The overall flowchart of the proposed \textbf{ADAPT}. ADAPT consists of a two-stage training paradigm and an adaptation module. \textbf{Stage (I):} an advertiser's bidding-period sequence, prefixed with a learnable CLS token and its category static profile, is encoded by a trainable bi-directional attention transformer into the hidden state, against which the advertiser's dynamic profile in a trainable memory bank is optimized by an InfoNCE loss. \textbf{Stage (II):} the same sequence is first prefixed with the frozen static profile; the common profile is then obtained by cross-attention with the individual profile as query and the memory bank as keys and values, while the private profile is obtained by encoding the individual profile with a multi-layer perceptron (MLP); both are prepended as prefixes and decorrelated by a correlation loss, and a DT with a linear head predicts the action under an MSE loss. \textbf{Adaptation:} for a new advertiser, the memory bank is fed as query, key, and value into the Stage (II) cross-attention and average pooled into a prototype common profile; the static profile, this prototype, and a zero private profile are prepended as prefixes and decoded by the frozen DT with a linear head into the action. For a new period, ADAPT reuses the Stage I encoding to obtain the hidden state and updates the advertiser's individual profile via an exponential moving average (EMA).}   
\label{fig:flowchart}
\vspace{-1.4em}
\end{figure*} 

Existing auto-bidding methods can generally be divided into two categories, \ie offline reinforcement learning (RL) methods and generative auto-bidding methods. Specifically, offline RL methods learn bidding policies purely from logged datasets, estimating value functions without further interaction with the environment. For example, IQL~\cite{kostrikovoffline} approximates the optimal value function via expectile regression and avoids querying out-of-distribution actions during training, thereby mitigating value overestimation and stabilizing policy learning. However, value estimation in long-horizon, sparse-reward bidding environments is often inaccurate, leading to unstable policy learning. To address this, generative auto-bidding methods reformulate auto-bidding as trajectory-conditioned action generation, bypassing value function learning and better capturing long-horizon dependencies~\cite{ajayconditional}. For instance, Decision Transformer (DT)~\cite{chen2021decision} employs a Transformer~\cite{vaswani2017attention} to model sequences of return-to-go (RTG), states, and actions, autoregressively generating bidding actions conditioned on a target return. Although generative methods have achieved remarkable success in auto-bidding, they typically train the model on historical trajectories to produce the current action, without personalizing to the underlying strategy of each advertiser. Meanwhile, profile-based methods have achieved remarkable success in domains such as recommendation~\cite{wang2025lettingo,zhang2026proex,zhang2024guided} and personalized generation and retrieval.~\cite{teevan2005personalizing,salemi2024lamp}. A user profile summarizes the style and preference of each user, thereby enabling personalization tailored to each individual. 

% Whether profiles can be leveraged to characterize the strategy of each advertiser and thus enable personalized bidding decisions remains a key yet open question.

Although profile-based methods have shown potential in other domains, directly applying them to auto-bidding still faces several challenges. First, existing profile-based methods mostly operate on discrete historical behaviors from which preference is easy to identify~\cite{zhang2024guided}, whereas an auto-bidding trajectory consists of continuous-valued states and actions from which the underlying strategy is difficult to extract. Second, profiles in domains such as recommendation are typically expressed in semantically explicit discrete forms that are easy to understand and inherently interpretable~\cite{wang2025lettingo}, whereas profiles in auto-bidding consist entirely of numerical information, making the encoded strategy difficult for the model to understand. Third, existing profile-based methods rely on side information or gradient updates to handle new users and profile drift~\cite{chen2018sequential}, whereas an advertiser exposes only its trajectory and online serving forbids retraining, forcing profile construction and updating to be training-free and trajectory-only.

% More importantly, how to build a complete auto-bidding framework that unifies profile extraction, effective utilization, and adaptive updating remains unexplored.

To address the above challenges, we propose a novel \underline{\textbf{A}}uto-bidding framework with \underline{\textbf{D}}isentangled \underline{\textbf{A}}dvertiser \underline{\textbf{P}}rofiles and \underline{\textbf{T}}raining-free adaptation, referred to as \textbf{ADAPT}. As illustrated in Figure \ref{fig:flowchart}, ADAPT is built upon a two-stage training paradigm followed by a training-free adaptation step. In the first stage, profile representation learning extracts pure static and dynamic profiles via contrastive learning over a shared advertiser memory bank, thereby addressing the first challenge. In the second stage, disentangled profile-based bidding applies a correlation loss to disentangle the dynamic profile into a common and a private profile for better profile understanding, which, together with the static profile, condition the final bidding decision, thereby addressing the second challenge. Finally, a training-free adaptation step cold-starts a new advertiser with a common profile distilled from existing advertisers and updates the dynamic profile of an existing advertiser without retraining, thereby addressing the third challenge. Compared with previous methods, the main contributions of our work are summarized as follows:
\begin{itemize}
    \item We propose \textbf{ADAPT}, an auto-bidding framework that learns disentangled advertiser profiles via two-stage training and maintains them via training-free adaptation, thereby delivering personalized bidding decisions.
    \item We show that ADAPT consistently outperforms strong baselines on a large-scale auto-bidding benchmark, with ablations confirming the contribution of each design.
    
    % \item We conduct \textbf{the first exploration }of parameter-efficient fine-tuning for LLM-based auto-bidding, by introducing the constraint-gated LoRA module that leverages task constraints as gating signals to dynamically select bidding strategies with minimal trainable parameters.
\end{itemize}

\section{Preliminaries}
\subsection{Problem Statement}
Modern online advertising platforms host thousands of advertisers that simultaneously contend for a limited supply of impressions. Within a given horizon, the traffic an advertiser faces forms a chronologically ordered trajectory of $D$ ad opportunities, and for the $i$-th opportunity ($i\in\{1,\ldots,D\}$) the advertiser responds with a bid $bid_i$. Each opportunity is allocated through an auction: the highest bidder wins the impression, receives its value $v_i$, and is charged a cost $c_i$ that, under the widely adopted second-price mechanism, is set by the runner-up bid. The advertiser aims to accumulate as much impression value as possible without exceeding a prescribed budget.

Beyond the budget, real-world campaigns frequently carry auxiliary requirements, a common one being the CPA constraint~\cite{he2021unified}. A competent auto-bidding strategy must therefore drive the advertiser's overall utility to its maximum while keeping every such constraint satisfied. This goal can be stated formally as:
\begin{equation}
\begin{aligned}
\max \quad & \sum_{i} o_i v_i \\
\text{s.t.} \quad & \sum_{i} o_i c_i \le B \\
& \frac{\sum_{i} c_{ij} o_i}{\sum_{i} p_{ij} o_i} \le C_j, \quad \forall j \\
& o_i \in \{0,1\}, \quad \forall i
\end{aligned}
\label{eq1}
\end{equation}
where $o_i$ denotes whether the advertiser wins the $i$-th impression, $v_i$ denotes its value, $B$ denotes the advertiser's budget, $c_{ij}$ denotes the cost incurred under constraint $j$ for impression $i$, $C_j$ denotes the advertiser-specified upper bound of the $j$-th constraint, and $p_{ij}$ denotes the corresponding performance indicator, \eg conversions. Prior work has shown that the constrained problem above admits a closed-form optimal bid~\cite{he2021unified}:
\begin{equation}
bid_i^* = \lambda_0 v_i + \sum_{j=1}^{J} \lambda_j p_{ij} C_j,
\label{eq2}
\end{equation}
where $bid_i^*$ denotes the optimal bid for the $i$-th impression, $\lambda_0$ denotes the optimal bidding parameter, $J$ denotes the number of constraints, and $\lambda_j$ denotes the Lagrange multiplier associated with the $j$-th constraint. Auto-bidding thus reduces to continually tuning these parameters so that the resulting bids stay optimal under multiple constraints.

\subsection{Sequential Decision for Auto-Bidding}
Because both competitor strategies and the distribution of arriving opportunities drift continuously, any fixed setting of the bidding parameters is soon rendered suboptimal. Optimal bidding therefore requires the parameters to be re-adjusted on the fly from real-time market feedback, which naturally casts auto-bidding as a sequential decision-making problem. We formalize this interaction through the following interconnected components, in which the model reads a state $s_t$, selects an action $a_t$, receives an immediate reward $r_t$ as feedback, and optimizes toward the cumulative RTG $R_t$:
\begin{itemize}
    \item State $s_t$: a feature vector summarizing the bidding environment at timestep $t$, such as the remaining budget, the remaining time, and historical bidding information.
    \item Action $a_t$: the action applied at timestep $t$, written as
    $a_t=(a^{\lambda_0}_t,a^{\lambda_1}_t,\ldots,a^{\lambda_J}_t)$.
    \item Reward $r_t$: the total value of the impressions won within timestep $t$. With $D_t$ candidate impressions in this period, it is defined as $r_t = \sum_{i=1}^{D_t} o_i v_i$, where $D_t$ is the number of ad opportunities at timestep $t$.
    \item RTG $R_t$: the reward accumulated from timestep $t$ to the terminal timestep $T$, defined as $R_t = \sum_{n=t}^{T} r_{n}$.
\end{itemize}
Under these definitions, the whole bidding procedure is represented as a trajectory $w = (R_0, s_0, a_0, \ldots, R_{T-1}, s_{T-1}, a_{T-1})$.

\section{Methodology}
\subsection{Motivation}
Generative auto-bidding methods have advanced in optimizing bids under various constraint values~\cite{li2026lbm,li2025gas,cai2026strategy}, but numerical constraints such as CPA cannot fully capture advertiser intent. Even the same CPA target may imply different bidding preferences across advertisers, highlighting the need for personalization. By modeling advertiser-specific strategies beyond constraint values, personalized auto-bidding can better align decisions with individual objectives, yet remains largely underexplored. Meanwhile, profile-based methods have shown strong results in other domains~\cite{wang2025lettingo,salemi2024lamp}. In recommendation systems, for instance, user profiles are built to capture individual preferences, allowing the system to provide items that better match each user’s interests and needs~\cite{zhang2024guided}. However, auto-bidding differs fundamentally from recommendation in both data structure and optimization objective. As a result, directly transferring existing profile-based methods to auto-bidding often fails to deliver the expected gains. 

More specifically, applying profile-based methods to auto-bidding faces several challenges. First, unlike recommendation profiles built from discrete user-item interactions, advertiser profiles must be distilled from historical bidding dynamics, where bids, costs, conversions, and constraints interact over time. This raises the question: \textit{how can we effectively extract multi-aspect profiles from an advertiser’s bidding history?} Second, recommendation profiles often carry explicit semantic information, such as item attributes or category labels, which helps the model interpret them. In auto-bidding, however, bidding histories and advertiser categories usually lack clear semantics, leaving the extracted profile as a dense vector whose shared and personalized components are hard to distinguish. This raises the question: \textit{how can the decision model better understand and use the profile?} Third, profile updates are more costly in auto-bidding. New advertisers require cold-start profiles, existing profiles must evolve as delivery continues, and retraining the model for every update is impractical. This raises the question:\textit{how can advertiser profiles be updated efficiently without retraining?} These challenges raise a key yet underexplored question: 
\begin{center}
	\textit{How can an auto-bidding framework systematically extract, exploit, and update advertiser profiles to enable personalized bidding decisions?}
\end{center}
 To this end, we propose \textbf{ADAPT}, which consists of a two-stage training paradigm and an adaptation module. During training, Stage 1 (Section~\ref{stage 1}) achieves effective profile extraction, while Stage 2 (Section~\ref{stage 2}) achieves effective profile exploitation and improves the model's understanding of the profile. The adaptation module (Section~\ref{adaptation}) then efficiently updates profiles whenever a new advertiser arrives or a new delivery period begins.
\subsection{Training}
We organize the training process into two stages. Stage 1 learns clean and stable advertiser profiles, while stage 2 uses the learned profiles to guide personalized bidding decisions. By assigning each stage a distinct objective, the training process prevents profiles from being overly shaped by action labels while ensuring that they are effectively exploited for decision making.

\subsubsection{Stage 1: Profile Representation Learning}
\label{stage 1}
Existing profile-based methods usually construct profiles by aggregating observed behaviors, which works well when user preferences are expressed through discrete and semantically meaningful interactions. In auto-bidding, however, an advertiser’s bidding preference cannot be fully described by a single type of information. On the one hand, the advertiser category provides useful static prior knowledge about its business domain. On the other hand, the advertiser’s actual bidding strategy is implicitly reflected in historical bidding dynamics, where returns, states, actions, and advertiser information interact over time. Therefore, we represent each advertiser from both static and dynamic perspectives: the static profile captures category-level prior information, while the dynamic profile is learned from bidding history to characterize advertiser-specific strategies.

Specifically, we divide the advertiser profile into a static profile and a dynamic profile. In this work, the static profile represents the advertiser category. For all advertiser categories, we build a trainable embedding table, denoted as $\bm{S}=\{\bm{s_0},\bm{s_1},\dots,\bm{s_{C-1}}\} \in \mathbb{R}^{C\times d}$, where $C$ is the number of categories and $d$ is the embedding dimension. The dynamic profile captures the personalized bidding strategy extracted from each advertiser’s historical delivery periods. We similarly initialize dynamic profiles with a trainable embedding table, denoted as $
\bm{P}=\{\bm{p_0},\bm{p_1},\dots,\bm{p_{N-1}}\} \in \mathbb{R}^{N\times d}$, where $N$ is the number of advertisers, and refer to it as the advertiser memory bank. The dynamic profile of advertiser $i$, denoted as $\bm{p}_i$, is referred to as the individual profile. For advertiser $i$ with category $c$ and we sample a fixed-length segment of length $K$ from its historical trajectory $w$, we first use a lightweight embedder to transform the trajectory into a sequence of trajectory embeddings $\bm{e}\in\mathbb{R}^{3K\times d}$. Since the goal is to extract the advertiser’s strategy information from the entire trajectory rather than to predict the next action, each position should be allowed to attend to both past and future contexts. Therefore, we adopt a bidirectional attention Transformer as the encoder $\mathcal{T}(\cdot)$. To construct the input sequence, we prepend two types of prefix tokens before the trajectory embeddings. The first is the corresponding static profile, which provides category-level strategy information for the advertiser. The second is a learnable cls token $\bm{cls}\in\mathbb{R}^d$, inspired by ViT~\cite{dosovitskiy2021an}, which is used to aggregate the strategy information of the entire trajectory. Positional embeddings are then added to the trajectory tokens to encode their temporal order. Finally, the input embeddings $\bm{x}$ to the encoder are formulated as:
\begin{equation}
    \bm{x}=[\bm{cls},\bm{s_c},\bm{e}],\ \ \ \ \bm{x}\in\mathbb{R}^{(2+3K)\times d}.
\end{equation}
After that, we feed the input embeddings into the encoder. The encoded feature corresponding to the cls token is then passed through layer normalization $\text{LayerNorm}(\cdot)$. This process is formulated as:
\begin{equation}
    \bm{h}=\text{LayerNorm}(\mathcal{T}(\bm{x})_0), \ \ \ \ \bm{h}\in\mathbb{R}^d,
\end{equation}
where $\bm{h}$ denotes the hidden state of the cls token. For the input trajectory of advertiser $i$, we take the corresponding memory entry $\bm{p}_i$ from $\bm{P}$ as its individual profile. Finally, we train Stage 1 with an InfoNCE-style contrastive objective, where $\bm{p}_i$ serves as the positive sample for $\bm{h}$ and while the profiles of other advertisers serve as negatives. The optimization process is formulated as:
\begin{equation}
    \mathcal{L}_\text{inf} 
    = -
    \log
    \frac{
    \exp(\bm{h}^{\top}\text{LayerNorm}(\bm{p}_{i}) /\tau)
    }{
    \sum_{j=1}^{N} \exp(\bm{h}^{\top} \text{LayerNorm}(\bm{p}_j) /\tau)
    },
\label{loss inf}
\end{equation}
where $\tau$ denotes the temperature coefficient. By optimizing Eq.(\ref{loss inf}), it learns advertiser-specific dynamic profiles that capture personalized bidding strategies and remain discriminative across advertisers. Overall, the profile representation learning stage extracts strategy-level information from bidding histories and form advertiser-specific dynamic profiles. It also learns category-aware static profiles and a strategy-aware trajectory encoder. As a result, stage 1 produces clean and informative advertiser profiles, addressing the first limitation.
\subsubsection{Stage 2: Disentangled Profile-based Bidding}
\label{stage 2}
Stage 1 produces clean and informative advertiser profiles from bidding histories and advertiser categories, but effectively exploiting these profiles remains non-trivial. Unlike recommendation profiles, which often come with explicit semantics such as item categories, textual descriptions, or user interests, auto-bidding profiles are continuous numerical tokens~\cite{zhang2024guided,wang2025lettingo}. In particular, a dynamic profile may encode both strategy patterns shared across advertisers and preferences specific to an individual advertiser. Directly using such dense profiles forces the decision model to interpret them implicitly, which can lead to insufficient profile utilization and redundant representations.

Based on the above analysis, it is necessary to transform the dense dynamic profile into semantically clearer components before using it for decision making. To do this, we further disentangle the dynamic profile into a common profile and a private profile, so that the decision model can better exploit the strategy information encoded in advertiser profiles. Specifically, the dynamic profile may contain strategy patterns that are commonly shared across advertisers. For example, most advertisers aim to obtain more conversions while satisfying CPA constraints, and such common bidding principles are useful for helping the model understand the general objective of auto-bidding. However, these shared strategies should not be represented by a single identical vector for all advertisers. Different advertisers may emphasize or express the same common strategy in different ways, depending on their own bidding histories and business preferences. Simply averaging all advertiser profiles or using a fixed shared profile would ignore such advertiser-dependent variations, weakening personalization and potentially introducing irrelevant information. Therefore, we use a cross-attention mechanism to extract the common profile. Given the individual profile of advertiser $i$, denoted as $\bm{p}_i$, we use it as the query and take the advertiser memory bank $\bm{P}$ as the keys and values. This process is formulated as:
\begin{equation}
    \bm{p}_\text{com}=\text{CrossAttn}(Q=\bm{p}_i,K=\bm{P},V=\bm{P}), \ \ \ \ \bm{p}_\text{com}\in\mathbb{R}^d,
\label{common}
\end{equation}
where $\bm{p}_\text{com}$ denotes the common profile, $Q$, $K$, and $V$ represent the query, key, and value embeddings, and $\text{CrossAttn}(Q,K,V)$  denotes a cross-attention layer. By using Eq.~(\ref{common}), the common profile extracts shared bidding patterns in an advertiser-aware manner, rather than relying on a fixed shared representation for all advertisers.

Similarly, different advertisers also have their private strategies beyond the shared ones. These strategies arise from advertiser-specific business goals, budget conditions, or market environments. Without modeling such private strategies, the decision model may overlook individual differences and produce overly homogeneous bidding decisions. Therefore, we extract the private profile directly from the individual profile with an MLP. This process is formulated as:
\begin{equation}
    \bm{p}_{\text{pri}} = \text{MLP}(\bm{p}_i), \ \ \ \ \bm{p}_\text{pri}\in\mathbb{R}^d,
\label{private}
\end{equation}
where $\bm{p}_\text{pri}$ denotes the private profile. By using Eq.~(\ref{private}), the private profile captures advertiser-specific preferences beyond shared bidding patterns. However, extracting the common and private profiles with the above modules does not guarantee that they are semantically separated. The two profiles may still contain overlapping or redundant information, causing the intended disentanglement to collapse into two similar representations. To avoid this, we introduce a correlation loss to encourage the common and private profiles to capture complementary information. Specifically, for each sample, we first center the common and private profiles along the hidden dimension, and then penalize the absolute cosine similarity between the centered vectors. The loss is formulated as:
\begin{equation}
\mathcal{L}_{\mathrm{cor}}
=
\frac{1}{B}
\sum_{z=1}^{B}
\frac{
\left|\mathrm{Cov}\left(\bm{p}_{\mathrm{com},z}, \bm{p}_{\mathrm{pri},z}\right)\right|
}{
\sqrt{\mathrm{Var}\left(\bm{p}_{\mathrm{com},z}\right)}
\sqrt{\mathrm{Var}\left(\bm{p}_{\mathrm{pri},z}\right)}
}
, \ \ \ \ \bm{p}_{\mathrm{com},z},\bm{p}_{\mathrm{pri},z}\in\mathbb{R}^d,
\label{corloss}
\end{equation}
 where $B$ denotes the batch size, $\bm{p}_{\mathrm{com},z}$ and $\bm{p}_{\mathrm{pri},z}$ denote the common and private profiles of the $z$-th sample respectively, $\mathrm{Cov}(\cdot,\cdot)$ and $\mathrm{Var}(\cdot)$ denote the covariance and variance computed across the $d$ hidden dimensions within each sample. By optimizing Eq.~(\ref{corloss}), ADAPT encourages the common and private profiles to be semantically separated, reducing redundancy and improving profile utilization. Next, for advertiser $i$ with category $c$, we sample a fixed-length segment of length $K$ from its historical trajectory $w$, and prepend the static, common, and private profiles to the trajectory embeddings $\bm{e}$:
 \begin{equation}
    \bm{y}=[\bm{s}_c,\bm{p}_{\mathrm{com}},\bm{p}_{\mathrm{pri}},\bm{e}],
    \quad
    \bm{y}\in\mathbb{R}^{(3+3K)	\times d}.
\end{equation}
The sequence $\bm{y}$ is encoded by a causal transformer as $\bm{H}=\mathrm{Backbone}(\bm{y})$, where $\bm{H}\in\mathbb{R}^{(3+3K)\times d}$. Then, the hidden state of each state token, denoted as $\bm{h}_t^s$, is linearly projected to predict the action, i.e., $\hat{a}_t=\mathrm{Linear}(\bm{h}_t^s)$. We use the mean squared error (MSE) loss to optimize action prediction:
\begin{equation}
    \mathcal{L}_{\mathrm{mse}}
    =
    \frac{1}{K}
    \sum_{t=0}^{K-1}
    \left\|a_t-\hat{a}_t
\right\|^2.
\end{equation}
The overall training objective of stage 2 is:
\begin{equation}
    \mathcal{L}
    =
    \mathcal{L}_{\mathrm{mse}}
    +
    \alpha*\mathcal{L}_{\mathrm{cor}},
\end{equation}
where $\alpha$ denotes the weight factor. Overall, stage 2 enables to exploit advertiser profiles through semantically clearer common and private components, thereby addressing the second limitation.

\subsection{Adaptation}
\label{adaptation}
Through the two-stage training process, ADAPT learns clean advertiser profiles and effectively exploits them for personalized bidding. However, real-world advertising environments are dynamic: existing advertisers continuously generate new delivery periods, while new advertisers may enter the system at any time. Without profile updates, existing profiles may become outdated and fail to reflect recent bidding strategies; without cold-start profiles, new advertisers cannot receive personalized decisions at the beginning of delivery. Therefore, efficient profile updating and cold-start construction are necessary for practical auto-bidding systems. To address this, we design a training-free adaptation mechanism. It updates the individual profile of an existing advertiser using newly observed delivery periods, and provides suitable strategy guidance for a new advertiser in a zero-shot manner when no learned individual profile is available.\\
\textbf{New period.}~~For existing advertisers, new periods may indicate recent strategy changes, making profile updating necessary. Retraining the profile extractor or fine-tuning the memory bank is a straightforward solution, but it is too costly for online advertising systems. Therefore , we update profiles in a training-free manner. Since the Stage 1 transformer has learned to summarize strategy information, we freeze it and encode the new period trajectory embedding $\bm{x}_\text{new}=\{\bm{cls},\bm{s}_c,\bm{e}_\text{new}\}$ of advertiser $i$ with category $c$: $\bm{h}_{\mathrm{new}}=\mathrm{LayerNorm}\left(\mathcal{T}(\bm{x}_{\mathrm{new}})_0\right)$. We then update its individual profile \(\bm{p}_i\) in the memory bank using exponential moving average (EMA):
\begin{equation}
    \bm{p}_i
    \leftarrow
    \lambda \bm{p}_i
    +
    (1-\lambda)\bm{h}_{\mathrm{new}},
\label{new period}
\end{equation}
where $\lambda$ controls the update strength. Eq.~(\ref{new period}) enables the individual profile to adapt to new bidding strategies while preserving historical information, without requiring retraining.\\
\textbf{New advertiser.}~~New advertisers may enter the bidding platform at any time, but they have no historical trajectories during training and thus no individual profiles in the memory bank. Retraining a profile for each new advertiser is inefficient, so we provide zero-shot strategy guidance using available profile information. Specifically, for a new advertiser $j$ with category $c$, we use its category static profile $\bm{s}_c$, average the common profiles of existing advertisers as prototype common profile $\bar{\bm{p}}_{\mathrm{com}}$, and set the private profile to a zero vector since no advertiser-specific strategy is available. The inference input is formulated as:
\begin{equation}
 \bm{y}=[\bm{s}_c,\bar{\bm{p}}_{\mathrm{com}},\bm{p}^0_\text{pri},\bm{e}], \ \ \  \ \bar{\bm{p}}_{\mathrm{com}},\bm{p}^0_\text{pri}\in\mathbb{R}^d,
 \label{new ad}
\end{equation}
where $\bm{p}^0_\text{pri}$ denotes the zero private profile. By using Eq.~(\ref{new ad}), ADAPT guides bidding decisions for new advertisers in a zero-shot manner, without requiring any additional training or model modification. Overall, the adaptation module enables efficient profile updating and zero-shot guidance without retraining, thereby addressing the third limitation.

Finally, ADAPT unifies profile extraction, exploitation, and updating into a complete advertiser profile modeling framework, addressing all three limitations discussed above.
\section{Experiments}
\subsection{Experimental Details}
\textbf{Datasets.}~~We conduct experiments on AuctionNet~\cite{su2024a}, a large-scale benchmark dataset for advertising bidding released by Alibaba. It provides two versions with different conversion sparsity levels, denoted as AuctionNet-dense and AuctionNet-sparse. Both versions contain records from 48 advertisers, and each advertising period involves over 500,000 impression opportunities. Additional dataset statistics are reported in Appendix~\ref{dataset}.\\
\textbf{Comparison methods.}~~The comparison methods include four offline RL methods (\ie USCB~\cite{he2021unified}, BCQ~\cite{fujimoto2019off}, CQL~\cite{kumar2020conservative}, IQL~\cite{kostrikovoffline}), and five generative auto-bidding methods(\ie DiffBid~\cite{guo2024generative}, DT~\cite{chen2021decision}, CDT~\cite{liu2023constrained}, DT-score~\cite{li2025gas}, GAS~\cite{li2025gas}).\\
\textbf{Implementation details.}~~Our method is implemented with a two-stage architecture. Stage 1 employs a bidirectional transformer with 8 layers to encode advertiser profiles via contrastive learning, while stage 2 uses a causal Transformer with 6 layers for sequential bidding decisions. The trajectory window length is $K{=}10$, and the correlation loss weight is set to $\alpha{=}1$. In the adaptation stage, the EMA coefficient is $\lambda{=}0.99$. Further implementation details are provided in Appendix~\ref{implementation}.\\
\textbf{Evaluation metrics.}~~We adopt two metrics for evaluation:
\begin{itemize}
    \item Conversions: the number of conversions achieved in a delivery period without considering additional constraints, computed as $\sum_i o_i v_i$.
    \item Score: the constraint-aware value under the CPA requirement. Specifically, we apply a penalty term $\min\left\{\left(\frac{1}{\text{ratio}}\right)^2,1\right\}$, where $\text{ratio}$ denotes the ratio of the achieved CPA to the target CPA. The final score is calculated as $\sum_i o_i v_i \times \text{penalty}$.
\end{itemize}
\begin{table*}[t]
\caption{Score comparison between our proposed method and previous baselines on the AuctionNet benchmark under different budget scales. Bold indicates the best performance, and underlined values denote the second-best performance.}
\vspace{-0.7em} 
\label{tab:1}
\centering

\setlength{\tabcolsep}{6pt}

\resizebox{\textwidth}{!}{
\begin{tabular}{l|c|cccccccccc|c}
\toprule
\textbf{Dataset} & \textbf{Budget} & \textbf{USCB} & \textbf{BCQ} & \textbf{CQL} & \textbf{IQL} & \textbf{DiffBid} & \textbf{DT} & \textbf{CDT} &\textbf{DT-score}& \textbf{GAS}& \textbf{ADAPT} & \textbf{Improve} \\
\midrule

\multirow{5}{*}{AuctionNet-dense}
& 50\%  & 86 & 190 & 113 & 164 & 54 &191  &174 &178&\underline{193}& \textbf{210}&8.81\%  \\
& 75\%  & 135 & 259 &139 & 232 & 100 &265  &242 &268&\underline{287} &\textbf{289} &0.70\%  \\
& 100\% & 157 & 321 &171 & 281 &152  &329  &326 &334&\underline{359}& \textbf{367}& 2.23\% \\
& 125\% & 220 & 379 &201 & 355 &193  &396  &378 &395&\underline{409} &\textbf{420} & 2.69\% \\
& 150\% & 281 & 429 &238  &401 &234  &450  &433 &441&\underline{461} &\textbf{469}& 1.74\% \\

\midrule

\multirow{5}{*}{AuctionNet-sparse}
& 50\%  &11.5  &17.7  &12.8  &16.5  &9.9  &14.8  &11.2 &17.7&\underline{18.4} &\textbf{18.7}& 1.60\%\\
& 75\%  &14.9  &24.6  &16.7  &22.1  &15.4  &22.9  &18.0 &25.9&\underline{27.5}&\textbf{29.5} &7.27\% \\
& 100\% &17.5  &31.1  &22.2  &30.0  &19.5  &29.6  &31.2 &33.2&\underline{36.1}&\textbf{37.1} & 2.77\%\\
& 125\% &26.7  &34.2  &28.6  &37.1  &25.3  &34.3  &31.7 &39.6&\underline{40.0}&\textbf{42.8} &7.00\%\\
& 150\% &31.3  &37.9  &35.8  &43.1  &30.8  &44.5  &39.1 &44.7&\underline{46.5} & \textbf{47.4}&1.94\%\\

\bottomrule
\end{tabular}
}
\vspace{-1em} 
\end{table*}

\subsection{Main Results}
We report the experimental results of all methods on the AuctionNet benchmark, including its two variants, AuctionNet-dense and AuctionNet-sparse. Table \ref{tab:1} compares our proposed method with previous auto-bidding methods.\\
\textbf{Comparison to baselines.}~~ADAPT achieves the best results across all budget scales on both variants. For instance, under the 100\% budget setting, ADAPT improves the score by 2.23\% and 2.77\% over the strongest baseline (\ie GAS) on AuctionNet-dense and AuctionNet-sparse, respectively, and by 141.45\% and 90.26\% over the weakest baseline (\ie DiffBid). The improvements mainly come from the advertiser profile modeling in ADAPT, which extracts strategy-level information from historical bidding trajectories and uses disentangled common and private profiles to provide personalized bidding guidance. This is especially helpful in sparse scenarios, where limited conversion feedback makes direct policy learning more difficult.
\begin{table*}[t]
\centering
\small
\begin{minipage}[t]{0.48\textwidth}
\centering
\caption{Effectiveness of advertiser profiles.}
\label{tab:ablation_profiles}
\vspace{-0.7em}
\begin{tabular}{lcc}
\toprule
\textbf{Method} 
& \textbf{Score $\uparrow$}
& \textbf{Conv. $\uparrow$}\\
\midrule
w/o Any Prefix (DT) & 329 & 364 \\
w/o Static Profile & 348 {\scriptsize (+5.78\%)} & 371 {\scriptsize (+1.92\%)} \\
w/o Dynamic Profile & 334 {\scriptsize (+1.52\%)} & 379 {\scriptsize (+4.12\%)} \\
ADAPT & \textbf{367} {\scriptsize (+11.55\%)} & \textbf{405} {\scriptsize (+11.26\%)} \\
\bottomrule
\end{tabular}
\end{minipage}
\hfill
\begin{minipage}[t]{0.48\textwidth}
\centering
\caption{Effectiveness of disentanglement.}
\label{tab:ablation_disentanglement}
\vspace{-0.7em}
\begin{tabular}{lcc}
\toprule
\textbf{Method} 
& \textbf{Score $\uparrow$}
& \textbf{Conv. $\uparrow$}\\
\midrule
w/o Disentanglement & 344 & 371  \\
w/o Private Profile & 348 {\scriptsize (+1.16\%)} & 379 {\scriptsize (+2.16\%)}\\
w/o Common Profile & 358 {\scriptsize (+4.07\%)} & 382 {\scriptsize (+2.96\%)} \\
ADAPT & \textbf{367} {\scriptsize (+6.69\%)} & \textbf{405} {\scriptsize (+9.16\%)} \\
\bottomrule
\end{tabular}
\end{minipage}
\vspace{-1.4em}
\end{table*}
\subsection{Ablation Studies}
We conduct a series of ablation studies to analyze the main components of ADAPT. Specifically, Table~2 evaluates the effectiveness of advertiser profiles; Table~3 studies the effectiveness of profile disentanglement; Table~4 examines the effectiveness of zero-shot new advertiser adaptation; and Table~5 investigates the effectiveness of training-free new period update. In addition, profile sensitivity analysis, hyperparameter studies, and profile visualization, are provided in Appendix~\ref{additional}.\\
\textbf{Effectiveness of advertiser profiles.}~~To evaluate the effectiveness of advertiser profiles, we compare ADAPT with variants that remove all profiles, use only the static profile, or use only the dynamic profile. As shown in Table~\ref{tab:ablation_profiles}, using only the dynamic profile improves  score and conversions by 5.78\% and 1.92\% over the no-profile variant, while using only the static profile improves them by 1.52\% and 4.12\%. When both profiles are used, ADAPT achieves the best performance, with 11.55\% and 11.26\% improvements in score and conversions. These results show that both profiles provide useful advertiser information, and their combination offers more effective guidance for bidding decisions.\\
\textbf{Effectiveness of profile disentanglement.}~~To verify the effectiveness of profile disentanglement, we compare ADAPT with variants that directly use the original individual profile, use only the common profile, or use only the private profile. As shown in Table~\ref{tab:ablation_disentanglement}, using only the common profile improves the score and conversions by 1.16\% and 2.16\% over the non-disentangled variant, while using only the private profile improves them by 4.07\% and 2.96\%, respectively. When both common and private profiles are used, ADAPT achieves the best results, improving the score and conversions by 6.69\% and 9.16\%. These results indicate that common and private profiles capture complementary strategy information, and their disentanglement enables more effective profile utilization.\\
\textbf{Zero-shot new advertiser adaptation.}~~To evaluate whether the adaptation module can generalize to unseen advertisers, we randomly select advertisers from different categories, remove all their data during the two-stage training, and test the model on these unseen advertisers. For fair comparison, DT is trained and evaluated under the same setting, and we report the averaged results over all selected advertisers in Table~
\ref{tab:prefix_ablation_new_adv}. DT achieves a score of 66, indicating that previous methods lack explicit strategy guidance for unseen advertisers. In contrast, adding the static profile improves the score and conversions by 6.60\% and 4.69\%, while adding the common profile further improves them by 7.55\% and 5.47\%. With the full zero-shot design, ADAPT achieves the best performance, improving the score and conversions by 26.42\% and 17.97\% over the zero-profile setting. These results show that the proposed profiles provide effective guidance for new advertisers and enable zero-shot generalization beyond previous auto-bidding methods.
\\
\textbf{Training-free new period update.}~~To verify whether ADAPT can adapt to newly observed bidding periods without training, we hold out 500 periods from the training set and do not use them during training. These periods are then treated as new periods to update advertiser profiles with the proposed EMA-based mechanism, while all model parameters remain frozen. As shown in Table~\ref{tab:adapt_results}, ADAPT improves the score and conversions by 5.39\% and 3.09\% over the variant without EMA update. These results indicate that the proposed update mechanism can incorporate recent bidding information into advertiser profiles without retraining, enabling efficient adaptation to new periods.
\\
\begin{table*}[t]
\centering
\small
\begin{minipage}[t]{0.48\textwidth}
\centering
\caption{Effectiveness of zero-shot new advertiser adaptation.}
\label{tab:prefix_ablation_new_adv}
\vspace{-0.5em}
\begin{tabular}{lcc}
\toprule
\textbf{Method} 
& \textbf{Score $\uparrow$}
& \textbf{Conv. $\uparrow$} \\
\midrule
DT & 66 & 73 \\
Zero Profiles & 106 & 128 \\
Static Profile & 113 {\scriptsize (+6.60\%)} & 134 {\scriptsize (+4.69\%)} \\
Common Profile & 114 {\scriptsize (+7.55\%)} & 135 {\scriptsize (+5.47\%)} \\
ADAPT (Zero-shot) & \textbf{134} {\scriptsize (+26.42\%)} & \textbf{151} {\scriptsize (+17.97\%)} \\
\bottomrule
\end{tabular}
\end{minipage}
\hfill
\begin{minipage}[t]{0.48\textwidth}
\centering
\caption{Effectiveness of training-free new period update.}
\label{tab:adapt_results}
\vspace{-0.5em}
\begin{tabular}{lcc}
\toprule
\textbf{Method} 
& \textbf{Score $\uparrow$}
& \textbf{Conv. $\uparrow$} \\
\midrule
w/o Update& 167 & 194 \\
ADAPT& \textbf{176} {\scriptsize (+5.39\%)} & \textbf{200} {\scriptsize (+3.09\%)} \\
\bottomrule
\end{tabular}
\end{minipage}
\vspace{-1.4em}
\end{table*}
\section{Conclusion}
In this paper, we propose \textbf{ADAPT}, a profile-based framework for personalized auto-bidding. Specifically, ADAPT first learns static and dynamic advertiser profiles from advertiser categories and historical bidding trajectories. It then disentangles the dynamic profile into common and private profiles to better capture shared bidding patterns and advertiser-specific preferences. Finally, ADAPT introduces a training-free adaptation mechanism to update profiles for new periods and provide zero-shot guidance for new advertisers. Experimental results demonstrate the effectiveness of ADAPT in improving bidding performance and enhancing adaptability in dynamic advertising environments.
\section*{AI Use Statement}
In this work, we used generative AI tools to assist with language polishing, grammar correction, readability improvement, and minor LaTeX formatting. We also used generative AI tools for limited code-related assistance, such as identifying syntax issues and improving code clarity. We did not use generative AI tools to generate datasets, design the core methodology, formulate mathematical claims, conduct experiments, or interpret experimental results. All AI-assisted text and code suggestions were carefully reviewed, revised, and verified by the authors. The authors take full responsibility for the final content of this work, including all claims, experimental results, and artifacts.
\section*{Acknowledgments}
This work was supported by Alibaba Group through Alibaba Research Intern Program.
\bibliography{iclr2027_conference}
\bibliographystyle{iclr2027_conference}

\clearpage
\appendix
\section{Appendix}
\subsection{Related Work}
\label{relate}
\textbf{Auto-bidding methods.}~~With the rapid growth of e-commerce platforms, auto-bidding for online advertising has become an important research topic. Early studies mainly relied on control-based methods~\cite{realtimechen,NIPS2017_da0d1111,yang2019bid,zhang2016feedback}, such as PID~\cite{realtimechen} and OnlineLP~\cite{NIPS2017_da0d1111}, which adjust bidding strategies through manually designed optimization or control rules. However, as auction environments become increasingly dynamic and the scale of advertising traffic continues to grow, these methods face significant challenges in scalability and adaptability. 

Benefiting from the rapid development of reinforcement learning and generative modeling, recent auto-bidding methods can be broadly grouped into two categories: traditional offline RL methods and generative auto-bidding methods. Traditional offline RL methods learn bidding policies by optimizing long-term rewards through value estimation and policy optimization~\cite{kumar2020conservative,kostrikovoffline,mou2022sustainable,wen2022cooperative}. For example, SORL~\cite{mou2022sustainable} combines offline pretraining with online RL to improve bidding stability, while MAAB~\cite{wen2022cooperative} models auto-bidding as a cooperative-competitive multi-agent optimization problem.

Generative auto-bidding methods instead formulate bidding as a trajectory generation or sequence modeling problem and directly produce bidding actions from historical trajectories or auxiliary information~\cite{chen2021decision,gao2025generative,lei2026generative,zhang2026generative}. For example, GAS~\cite{li2025gas} enhances generative bidding through post-training search, while DRIVE~\cite{cui2026drive} leverages distributional modeling and retrieval augmentation to improve decision quality. Some methods further incorporate planning mechanisms to guide current bidding decisions by modeling future dynamics~\cite{gao2025segb,guo2024generative,janner2022planning,li2025generative}. For instance, DiffBid~\cite{guo2024generative} models bidding behaviors through conditional diffusion processes, while SEGB~\cite{gao2025segb} employs self-evolved autoregressive diffusion for long-horizon bidding planning. Recently, LLM-based multi-modal methods have extended generative auto-bidding by leveraging the reasoning and language understanding capabilities of LLMs. These methods use textual information to model advertiser objectives and constraints while benefiting from the long-context reasoning ability of LLMs~\cite{lv2026decisionllm,li2026lbm,zhu2026role,cai2026strategy}. For example, LBM~\cite{li2026lbm} adopts a hierarchical reasoning-and-acting framework for bidding decision generation, SemBid~\cite{zhu2026role} investigates the role of language representations in improving auto-bidding performance, and SAGE~\cite{cai2026strategy} introduces a strategy-aware parameter-efficient adaptation framework for LLM-based auto-bidding.\\

\textbf{Profile-based personalization.}~~Profiles play an important role in personalized decision-making, as they summarize user preferences, behavioral patterns, and contextual characteristics into compact representations. They have been widely studied in recommendation systems, personalized generation, and personalized retrieval~\cite{wang2025lettingo,zhang2026proex,zhang2024guided,teevan2005personalizing,salemi2024lamp}. In recommendation, user profiles are commonly constructed from historical interactions or side information to better match users with items. Recent studies further leverage LLMs to generate more interpretable and semantically rich profiles. For example, LettinGo~\cite{wang2025lettingo} uses multiple LLMs to explore diverse user profiles and aligns profile generation with downstream recommendation performance through preference optimization. PALR~\cite{yang2023palr} incorporates user interaction histories into LLM-based recommendation and fine-tunes a large ranking model to select preferred items from retrieved candidates. GPG~\cite{zhang2024guided} generates natural-language user profiles from sparse and complex personal contexts, enabling LLMs to better capture user habits and preferences. ProEx~\cite{zhang2026proex} further constructs multiple profiles for each user or item to model different aspects of user intent and improve both discriminative and generative recommendation models.

Beyond recommendation, profiles have also been used for personalized generation and retrieval. In personalized search, user profiles or implicit behavioral histories, such as previous queries, browsing records, and clicked pages, are used to infer user intent and re-rank retrieval results. In personalized language generation, profiles provide user-specific context for generating outputs that better match individual preferences and writing styles~\cite{teevan2005personalizing,salemi2024lamp}. For instance, LaMP~\cite{salemi2024lamp} provides a benchmark for personalized language modeling and studies retrieval-augmented methods that select relevant profile items to condition model outputs. Prior work on personalized dialogue, review generation, and headline generation also shows that incorporating user or persona profiles can improve the relevance and consistency of generated content~\cite{teevan2005personalizing,zhang2018personalizing,ao2021pens}. Different from these domains, advertiser profiles in auto-bidding are usually derived from continuous bidding trajectories rather than explicit text or interaction semantics, making profile construction and utilization more challenging.

\subsection{Dataset Details}
\label{dataset}
\begin{table}[h]
\centering
\caption{Parameters of AuctionNet-dense and AuctionNet-Sparse.}
\label{tab:dataset}
\resizebox{0.6\columnwidth}{!}{
\begin{tabular}{lcc}
\toprule
\textbf{Parameters} & \textbf{AuctionNet-dense} & \textbf{AuctionNet-sparse} \\
\midrule
Trajectories & 479,376 & 479,376 \\
Delivery periods & 9,987 & 9,987 \\
timestep in a trajectory & 48 & 48 \\
RTG dimension & 1 & 1 \\
State dimension & 16 & 16 \\
Action dimension & 1 & 1 \\
Action range & [0, 493] & [0, 589] \\
Impression's value range & [0, 1] & [0, 1] \\
CPA range & [6, 12] & [60, 130] \\
Total conversion range & [0, 1512] & [0, 57] \\
\bottomrule
\end{tabular}
}
\end{table}
In our experiments, we adopt AuctionNet~\cite{su2024a}, a large-scale public advertising bidding benchmark proposed and constructed by Alibaba. Specifically, the dataset provides two variants with different conversion sparsity levels, referred to in this paper as AuctionNet-dense and AuctionNet-sparse. Each variant contains over 500000 impression opportunities from 48 advertisers across 9987 delivery periods. Detailed dataset statistics are shown in Table \ref{tab:dataset}. In addition, the state at each timestep in a trajectory is constructed by aggregating historical statistics and current budget-related information. Specifically, each dimension represents the following information:
\begin{description}[leftmargin=0pt,labelsep=0.5em,font=\normalfont\bfseries]
    \item[time\_left:] The number of remaining decision intervals in the current advertising delivery period.
    
    \item[budget\_left:] The remaining budget available for allocation in the current delivery period.
    
    \item[historical\_bid\_mean:] The average bid submitted by the advertiser over all previous timesteps.
    
    \item[last\_three\_bid\_mean:] The moving average bid over the last three timesteps.
    
    \item[historical\_LeastWinningCost\_mean:] The average market clearing price over all previous timesteps, \ie the minimum cost required to win an auction.
    
    \item[last\_three\_LeastWinningCost\_mean:] The average market clearing price over the last three timesteps.
    
    \item[historical\_pValues\_mean:] The average predicted conversion probability over all previous timesteps.
    
    \item[last\_three\_pValues\_mean:] The average predicted conversion probability over the last three timesteps.
    
    \item[current\_pValues\_mean:] The average predicted conversion probability of impression opportunities arriving at the current timestep.
    
    \item[historical\_conversion\_mean:] The average number of conversions achieved over all previous timesteps.
    
    \item[last\_three\_conversion\_mean:] The average number of conversions over the last three timesteps.
    
    \item[historical\_xi\_mean:] The historical average winning rate, defined as the ratio between the number of won auctions and the total number of participated auctions.
    
    \item[last\_three\_xi\_mean:] The average winning rate over the last three timesteps.
    
    \item[current\_pv\_num:] The total number of available impression opportunities at the current timestep.
    
    \item[last\_three\_pv\_num\_total:] The cumulative number of impression opportunities over the last three timesteps.
    
    \item[historical\_pv\_num\_total:] The cumulative number of impression opportunities over all previous timesteps.
\end{description}
\begin{table}[h]
\centering
\caption{Detailed hyperparameter settings for ADAPT.}
\label{tab:hyper}
\resizebox{0.4\columnwidth}{!}{
\begin{tabular}{lc}
\toprule
\textbf{Hyperparameter} & \textbf{Value} \\
\midrule
Layer of stage 1 transformer&8\\
Layer of stage 2 transformer&6\\
Head of stage 1 transformer&8\\
Head of stage 2 transformer&8\\
Batch size & 128  \\
Total training steps of stage 1 & 200000 \\
Total training steps of stage 2 & 400000 \\
Learning rate & 1e-5  \\
Weight decay& 1e-4 \\
Optimizer & AdamW  \\
Stage-1/2 trajectory length $K$ & 10 \\
Adaptation trajectory length $K$ & 48 \\
Episode length & 48 \\
\bottomrule
\end{tabular}
}
\end{table}
\subsection{Implementation Details}
\label{implementation}
Table \ref{tab:hyper} presents the detailed hyperparameter configurations of the proposed ADAPT method. Notably, the hyperparameter settings remain largely consistent across different benchmark variants, which further demonstrates the robustness of our method.
\subsection{Additional Ablation Studies and Visualizations}
\label{additional}
\begin{table}[t]
\centering
\small
\caption{Sensitivity analysis of advertiser profiles. At inference time we perturb the static and/or dynamic profile by replacing it with a zero vector (\emph{Zero}) or by randomly reassigning it to another advertiser (\emph{Shuffle}); \emph{Intact} denotes the unperturbed profile.}
\label{tab:profile_sensitivity}
\vspace{-0.5em}
\setlength{\tabcolsep}{6pt}
\begin{tabular}{cccccccc}
\toprule
\multicolumn{3}{c}{\textbf{Static Profile}}
& \multicolumn{3}{c}{\textbf{Dynamic Profile}}
& \multirow{2}{*}{\textbf{Score $\uparrow$}}
& \multirow{2}{*}{\textbf{Conv. $\uparrow$}} \\
\cmidrule(lr){1-3} \cmidrule(lr){4-6}
Intact & Zero & Shuffle & Intact & Zero & Shuffle & & \\
\midrule
           & \checkmark &            & \checkmark &            &            & 361 & 394 \\
           &            & \checkmark & \checkmark &            &            & 341 & 390 \\
\addlinespace[2pt]
\checkmark &            &            &            & \checkmark &            & 347 & 373 \\
\checkmark &            &            &            &            & \checkmark & 348 & 386 \\
\addlinespace[2pt]
           & \checkmark &            &            & \checkmark &            & 352 & 393 \\
           &            & \checkmark &            &            & \checkmark & 334 & 377 \\
\midrule
\checkmark &            &            & \checkmark &            &            & \textbf{367} & \textbf{405} \\
\bottomrule
\end{tabular}
\end{table}

\textbf{Profile sensitivity analysis.}~~To further verify whether the learned profiles contain meaningful advertiser information, we perturb the static and dynamic profiles at inference time and report the results in Table~\ref{tab:profile_sensitivity}. Specifically, we either replace a profile with a zero vector or randomly shuffle it across advertisers, while keeping the model parameters unchanged. The results show that perturbing either profile consistently degrades performance. For example, shuffling the static profile reduces the score from 367 to 341, while zeroing and shuffling the dynamic profile reduce the score to 347 and 348, respectively. When both profiles are shuffled, the score further drops to 334. These results indicate that the learned profiles are not redundant prefix tokens, but encode meaningful category-level and advertiser-specific strategy information for bidding decisions, further validating the effectiveness of our profile modeling design.\\
\begin{figure*}[t!] % [t] 表示将图片浮动到页面顶部
\centering 
\includegraphics[width=0.6\textwidth]{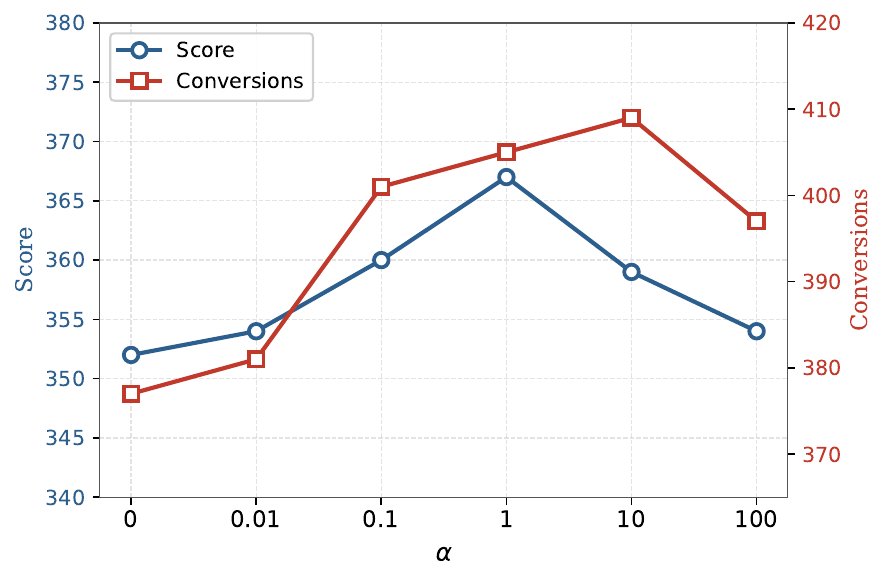} 
\vspace{-1.4em} 
\caption{Performance sensitivity to the correlation loss weight $\alpha$.}   
\label{fig:alpha_sensitivity}
\vspace{-1.4em}
\end{figure*} 
\noindent\textbf{Hyperparameter sensitivity analysis.}~~To analyze the effect of the correlation loss, we vary the loss weight $\alpha$ in $\mathcal{L}=\mathcal{L}_\text{mse}+\alpha*\mathcal{L}_\text{cor}$ and report the results in Figure~
\ref{fig:alpha_sensitivity}. When $\alpha=0$, the correlation loss is removed and the model achieves a score of 352. Introducing the correlation loss consistently improves performance, with the best score obtained at $\alpha=1$. In addition, ADAPT maintains strong performance across a reasonable range of $\alpha$, such as $0.1$, $1$, and $10$, indicating that the method is not overly sensitive to this hyperparameter. These results verify the effectiveness of the correlation loss and demonstrate the robustness of ADAPT to the choice of $\alpha$.\\
\begin{figure*}[t!] % [t] 表示将图片浮动到页面顶部
\centering 
\includegraphics[width=\textwidth]{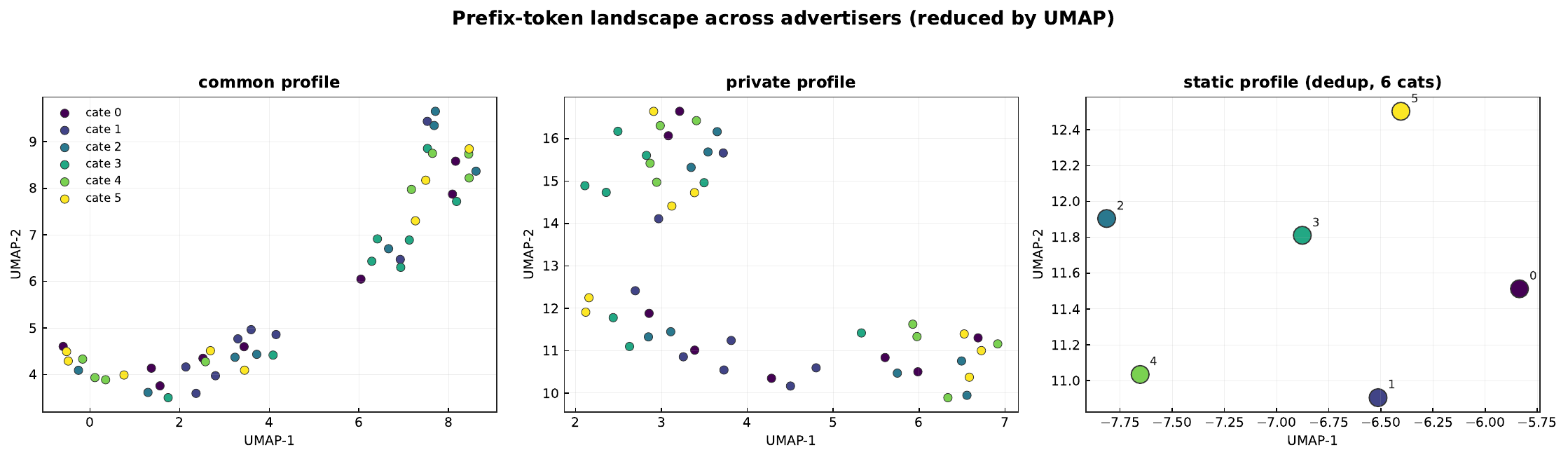} 
\vspace{-1.4em} 
\caption{UMAP visualization of learned advertiser profiles.}   
\label{fig:profile_scatter}
\vspace{-1.4em}
\end{figure*} 
\begin{figure*}[t!] % [t] 表示将图片浮动到页面顶部
\centering 
\includegraphics[width=\textwidth]{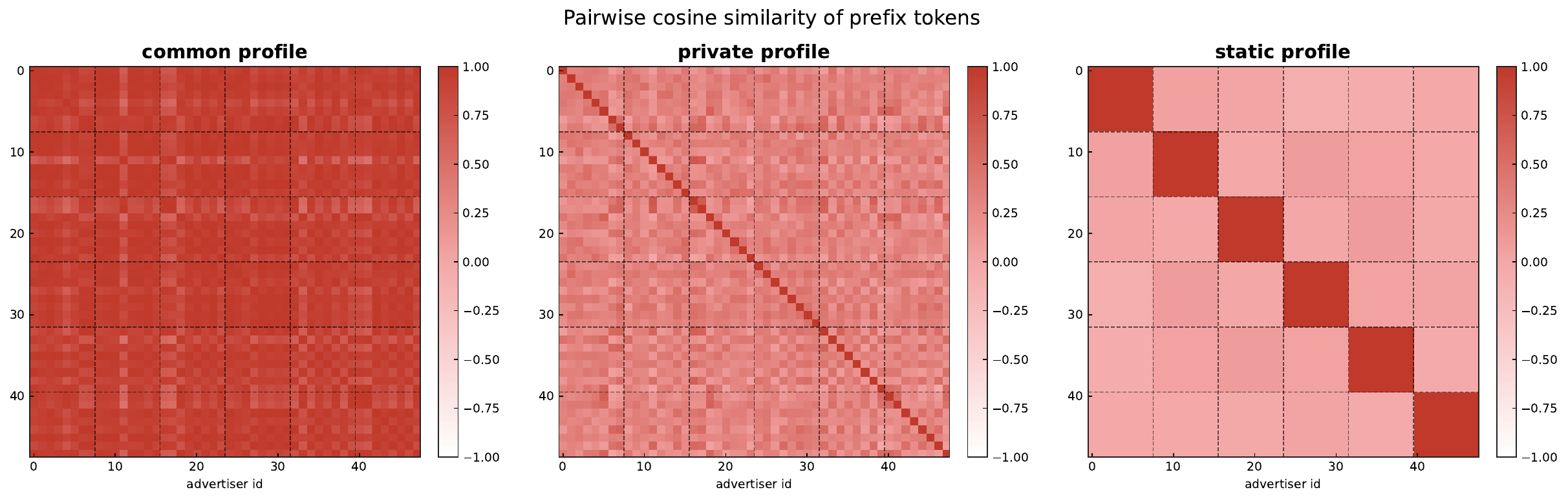} 
\vspace{-1.4em} 
\caption{Visualization of pairwise cosine similarity among learned advertiser profiles.}
\label{fig:profile_similarity}
\vspace{-1.4em}
\end{figure*} 
\textbf{Profile Visualization and Similarity Analysis.}~~To further examine whether the learned profiles are consistent with their expected semantics, we conduct visualization and similarity analysis for the static, common, and private profiles. Specifically, we first project the learned profiles into a two-dimensional space using UMAP, as shown in Figure~\ref{fig:profile_scatter}. We then visualize the pairwise cosine similarity between advertisers for each type of profile in Figure~\ref{fig:profile_similarity}. In addition, we report the mean similarity, intra-category similarity, and inter-category similarity in Table~\ref{tab:prefix_similarity}.

From Figure~\ref{fig:profile_scatter}, we observe that common profiles are relatively clustered, indicating that they capture shared bidding patterns across advertisers. In contrast, private profiles are more scattered, suggesting that they encode advertiser-specific strategy preferences. The static profiles form separated category-level representations and are clearly different from the dynamic profiles. Moreover, the common and private profiles occupy largely different regions, showing that the disentanglement process separates shared and individual strategy information.

Figure~\ref{fig:profile_similarity} further visualizes the pairwise similarity patterns. The common profiles show high similarity across advertisers, indicating that they represent general bidding knowledge without strong advertiser-specific preference. The private profiles exhibit high similarity mainly on the diagonal, while the similarities between different advertisers are much lower, confirming their personalized nature. For the static profiles, advertisers belonging to the same category share the same category-level representation and therefore show higher within-category similarity than cross-category similarity, indicating that static profiles encode category-specific information.

Table~\ref{tab:prefix_similarity} provides quantitative evidence for these observations. The common profiles have a high mean similarity of 0.9027, while the private profiles have a much lower mean similarity of 0.3218, confirming that common profiles capture shared strategies and private profiles preserve personalized information. Moreover, both common and private profiles show similar intra-category and inter-category similarities, indicating that they contain limited category-specific information after disentanglement. In contrast, the static profiles show high intra-category similarity and very low inter-category similarity, demonstrating their role in providing complementary category-level information. Overall, these results show that ADAPT learns meaningful profiles with distinct semantics, and the static, common, and private profiles are complementary and consistent with their expected roles.\\

\begin{table}[t]
\centering
\caption{Pairwise cosine similarity statistics of prefix profiles.}
\label{tab:prefix_similarity}
\begin{tabular}{lccc}
\toprule
Profile & Mean & Intra-category & Inter-category \\
\midrule
Common Profile  & 0.9027 & 0.8988 & 0.9034 \\
Private Profile& 0.3218 & 0.3158 & 0.3228 \\
Static Profile & 0.1532 & 1.0000 & 0.0050 \\
\bottomrule
\end{tabular}
\end{table}
% 需要宏包: caption (提供 \captionof)

% \begin{figure*}[t!]
% \centering
% \begin{minipage}[t]{0.46\textwidth}
% \centering
% \captionof{table}{Detailed hyperparameter settings for ADAPT.}
% \label{tab:hyper}
% \vspace{-0.5em}
% \footnotesize
% \begin{tabular}{lc}
% \toprule
% \textbf{Hyperparameter} & \textbf{Value} \\
% \midrule
% Layer of stage 1 transformer & 8 \\
% Layer of stage 2 transformer & 6 \\
% Head of stage 1 transformer & 8 \\
% Head of stage 2 transformer & 8 \\
% Batch size & 128 \\
% Total training steps of stage 1 & 200000 \\
% Total training steps of stage 2 & 400000 \\
% Learning rate & 1e-5 \\
% Weight decay & 1e-4 \\
% Optimizer & AdamW \\
% Stage-1/2 trajectory length $K$ & 10 \\
% Adaptation trajectory length $K$ & 48 \\
% Episode length & 48 \\
% \bottomrule
% \end{tabular}
% \end{minipage}
% \hfill
% \begin{minipage}[t]{0.50\textwidth}
% \centering
% \vspace{0pt}
% \includegraphics[width=\linewidth]{alpha_ablation.pdf}
% \vspace{-1.2em}
% \caption{Performance sensitivity to the correlation loss weight $\alpha$.}
% \label{alpha}
% \end{minipage}
% \vspace{-1.2em}
% \end{figure*}
\end{document}

%% file: math_commands.tex
\usepackage{amsmath,amsfonts,bm}

\def\eqref#1{equation~\ref{#1}}
\def\1{\bm{1}}

\DeclareMathAlphabet{\mathsfit}{\encodingdefault}{\sfdefault}{m}{sl}
\SetMathAlphabet{\mathsfit}{bold}{\encodingdefault}{\sfdefault}{bx}{n}